# A Tale of Two Animats: What does it take to have goals?

Larissa Albantakis

*Wisconsin Institute for Sleep and Consciousness*
*Department of Psychiatry, University of Wisconsin, Madison, WI, USA*
albantakis@wisc.edu

What does it take for a system, biological or not, to have goals? Here, this question is approached in the context of *in silico* artificial evolution. By examining the informational and causal properties of artificial organisms ("animats") controlled by small, adaptive neural networks (Markov Brains), this essay discusses necessary requirements for intrinsic information, autonomy, and meaning. The focus lies on comparing two types of Markov Brains that evolved in the same simple environment: one with purely feedforward connections between its elements, the other with an integrated set of elements that causally constrain each other. While both types of brains 'process' information about their environment and are equally fit, only the integrated one forms a causally autonomous entity above a background of external influences. This suggests that to assess whether goals are meaningful for a system itself, it is important to understand what the system *is*, rather than what it *does*.

## 0. Prequel

It was a dark and stormy night, when an experiment of artificial evolution was set into motion at the University of Wisconsin-Madison. Fifty independent populations of adaptive Markov Brains, each starting from a different pseudo-random seed, were released into a digital world full of dangers and rewards. Who would make it into the next generation? What would their neural networks look like after 60,000 generations of selection and mutation?

While electrical signals were flashing inside the computer, much like lightning on pre-historic earth, the scientist who, in god-like fashion, had designed the simulated universes and set the goals for survival, waited in suspense for the simulations to finish. What kind of creatures would emerge? ...

## I. Introduction

Life, from a physics point of view, is often pictured as a continuous struggle of thermodynamically open systems to maintain their complexity in the face of the second law of thermodynamics—the overall increase of entropy in our universe [1–3]. The 'goal' is survival. But is our universe like a game, in which organisms, species, or life as a whole increase their score by surviving? Is there a way to win? Does life have a chance if the 'goal' of the universe is a maximum entropy state ('death')?

Maybe there is an underlying law written into the fabrics of our universe that aligns the 'goal' of life with the 'goal' of the universe. Maybe 'information' is fundamental to discover it [4] (see also Carlo Rovelli's essay contribution). Maybe all there is are various gradients, oscillations, or fluctuations. In any case, looming behind these issues, another fundamental question lingers: What does it take for a system, biological or not, to have goals?

To approach this problem with minimal confounding factors, let us construct a universe from scratch: discrete, deterministic, and designed with a simple set of predefined, built-in rules for selection. This is easily done within the realm of *in silico* artificial evolution. One such world is shown in Fig. 1A (see also [5]). In this environment, the imposed goal is to categorize blocks of different sizes into those that have to be caught ('food') and those that have to be avoided ('danger'), limiting life to the essential. Nevertheless, this task requires temporal-spatial integration of sensor inputs and internal states (memory), to produce appropriate motor responses. Fitness is measured as the number of successfully caught and avoided blocks.

Let us then populate this simulated universe with 'animats', adaptive artificial organisms, equipped with evolvable Markov Brains [5,6]. Markov Brains are simple neural networks of generalized logic gates, whose input-output functions and connectivity are genetically encoded.

For simplicity, the Markov Brains considered here are constituted of binary, deterministic elements. Over the course of thousands of generations, the animats adapt to their task environment through cycles of fitness-based selection and (pseudo) random genetic mutation (Fig. 1B). One particularly simple block-categorization environment requires the animats to catch blocks of size 1 and avoid blocks of size 3 ("c1-a3") to increase their fitness.

*In silico* evolution experiments have the great advantage that they can easily be repeated many times, with different initial seeds. In this way, a larger portion of the 'fitness landscape', the solution space of the task environment, can be explored. In the simple c1-a3 environment, perfect solutions (100% fitness) were achieved at the end of 13 out of 50 evolution experiments starting from independent populations run for 60,000 generations. In the following we will take a look at the kind of creatures that evolved.

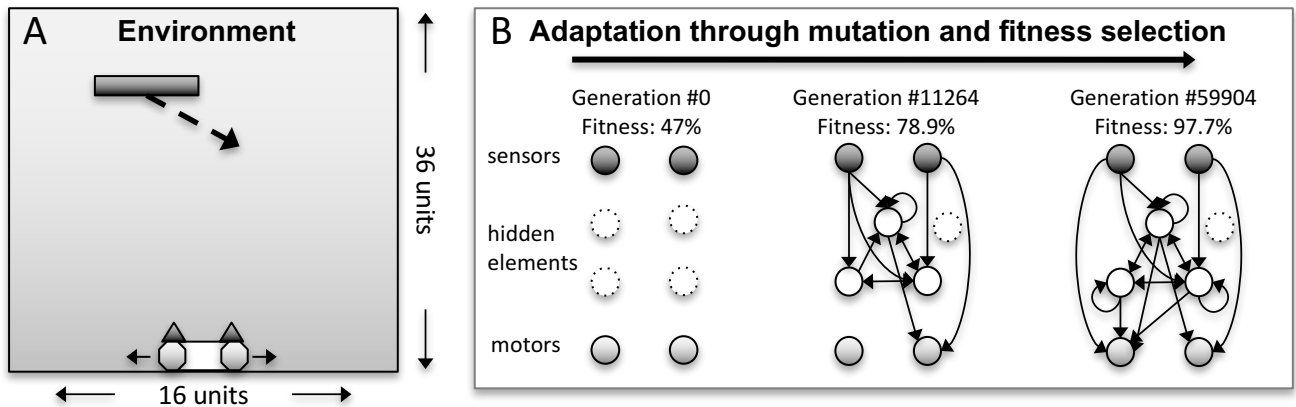

*Fig. 1:* *Artificial evolution of animats controlled by Markov Brains. (A) The animat is placed in a 16 by 36 world with periodic boundaries to the left and right. An animat's sensors are activated when a block is positioned above them, regardless of distance. Blocks of different sizes fall one at a time to the right or left. Animats can move to the left or right one unit per update. (B) An animat's Markov Brain is initialized without connections between elements and adapts to the task environment through fitness-based selection and probabilistic mutation. Adapted from* [7] *with permission.*

## II. Perfect fitness—goal achieved?

deterministic, consists of at most 8 elements, and we have perfect knowledge of its logic structure. While there is no single elegant equation that captures an animat's internal dynamics, we can still describe and predict the state of its elements, how it reacts to sensor inputs, and when it activates its motors, moment by moment, for as long as we want. Think of a Markov Brain as a finite cellular automaton with inputs and outputs. No mysteries.

On the other hand, we may still aim for a comprehensive, higher-level description of the animat's behavior. One straightforward strategy is to refer to the goal of the task: "the animat tries to catch blocks of size 1 and avoid blocks of size 3". This is, after all, the rule we implemented for fitness selection. It is the animat's one and

$R$ captures information about features of the environment encoded in the internal states of the Markov Brain beyond the information present in its sensors. Conditioning on the sensors discounts information that is directly copied from the environment at a particular time step. A simple camera would thus have zero representation, despite its capacity to make $> 10^7$ bit copies of the world.

For animats adapting to the block-catching task, relevant environmental features include whether blocks are small or large, move to the left or right, etc. Indeed, representation $R$ of these features increases, on average, over the course of evolution [6]. While this result implies that representation of environmental features, as defined above, is related to task fitness, the measure *R* itself does not capture whether or to what extent the identified representations actually play a *causal* role in determining an animat's behavior[1].

Machine-learning approaches, such as decoding, provide another way to identify whether and where information about environmental variables is present in the evolved Markov Brains. Classifiers are trained to predict environmental categories from brain states—a method now frequently applied to neuro-imaging data in the neurosciences [9,10]. Roughly, the better the prediction, the more information was available to the classifier. Just as for *R*, however, the fact that information about specific stimuli can be extracted from a brain's neural activity does not necessarily imply that the brain itself is 'using' this information [11].

What about our animats? As demonstrated in Fig. 2, the c1-a3 block-categorization task can be perfectly solved by animats with as few as 2 hidden elements. Their capacity for representation is thus bounded by 4 bits (2 hidden elements + 2 motors). Is that sufficient for a representation of the goal for survival? At least in principle, 4 binary categories could be 'encoded'. Yet, in practice, even a larger version of animats with higher capacity for representation (10 hidden elements) only achieved values on the order of *R* = 0.6 bits in a similar block-catching environment [6]. To solve this task, the animats thus do not seem to require much categorical information about the environment beyond their sensor inputs.

While this lack of representation in the animats may be due to their small size and the simplicity of the task, there is a more general problem with the type of information measures described above: the information that is quantified is, by definition, *extrinsic* information.

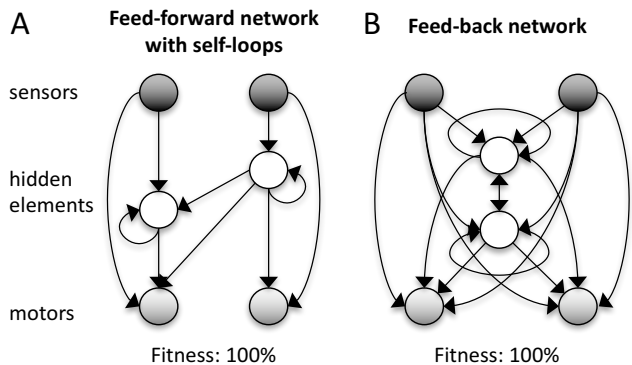

*Fig. 2: Example network architectures of evolved Markov Brains that achieved perfect fitness in the c1-a3 block-catching task. Adapted from* [5] *with permission.*

---

[1] Furthermore, representations of individual environmental features are typically distributed across many elements [6], and thus do no coincide with the Markov Brain's elementary (micro) logic components.

Any form of representation is ultimately a correlation measure between external and internal states, and requires that relevant environmental features are preselected and categorized by an independent observer (e.g. to obtain *E* in eq. 1, or to train the decoder). As a consequence, the information about the environment *represented* in the animat's Markov Brain is meaningful for the investigator. Whether it is causally relevant, let alone meaningful, for the animat is not addressed.[2]

## III. Intrinsic information

To be causally relevant, information must be physically instantiated. For every 'bit', there must be some mechanism that is in one of two (or several) possible states, and which state it is in must matter to other mechanisms. In other words, the state must be "a difference that makes a difference" [12,13].

More formally, a mechanism *M* has inputs that can influence it and outputs that are influenced by it. By being in a particular state *m*, *M* constrains the possible past states of its inputs[3], and the possible futures states of its outputs in a specific way. How much *M* in state *m* constrains its inputs can be measured by its *cause information* (*ci*); how much it constrains its outputs is captured by its *effect information* (*ei*) [13].

An animat's Markov Brain is a set of interconnected logic elements. A mechanism *M* inside the Markov Brain could be one of its binary logic elements, but can in principle also be a set of several such elements[4]. In discrete dynamical systems, such as the Markov Brains, with discrete updates and states, we can quantify the cause and effect information of a mechanism *M* in its current state $m_t$ within system *Z* as the difference *D* between the constrained and unconstrained probability distributions over *Z*'s past and future states [13]:

$$ci(M = m_t) = D\big(p(z_{t-1}|m_t), p(z_{t-1})\big) \quad (2)$$

$$ei(M = m_t) = D\big(p(z_{t+1}|m_t), p(z_{t+1})\big) \quad (3)$$

where $z_{t-1}$ are all possible past states of *Z* one update ago, and $z_{t+1}$ all possible future states of *Z* at the next update. For $p(z_{t-1})$, we assume a uniform (maximum entropy) distribution, which corresponds to perturbing *Z* into all possible states with equal likelihood. Using such systematic perturbations makes it possible to distinguish observed correlations from causal relations [14][5]. By evaluating a causal relationship in all possible contexts (all system states), we can obtain an objective measure of its specificity ("Does A always lead to B, or just sometimes?") [13,15]. Likewise, we take $p(z_{t+1})$ to be the distribution obtained by providing independent, maximum entropy inputs to each of the system's elements [13]. In this way, Eqn. 2 and 3 measure the causal specificity with which mechanism $M$ in state $m_t$ constrains the system's past and future states.

---

[2] Note that this holds, even if we could evaluate the correlation between internal and external variables in an observer-independent manner, except then the correlations might not even be meaningful for the investigator.

[3] If *M* would not constrain its inputs, its state would just be a source of noise entering the system, not causal information.

[4] Sets of elements can constrain their joint inputs and outputs in a way that is irreducible to the constraints of their constituent elements taken individually [13]. The irreducible cause-effect information of a set of elements can be quantified similarly to Eqn. 2-3, by partitioning the set and measuring the distance between $p(z_{t\pm1}|m_t)$ and the distributions of the partitioned set.

A system can only 'process' information to the extent that it has mechanisms to do so. All causally relevant information within a system $Z$ is contained in the system's *cause-effect structure*, the set of all its mechanisms, and their cause and effect distributions $p(z_{t-1}|m_t)$ and $p(z_{t+1}|m_t)$. The cause-effect structure of a system in a state specifies the information *intrinsic* to the system, as opposed to correlations between internal and external variables. If the goals that we ascribe to a system are indeed meaningful from the intrinsic perspective of the system, they must be intrinsic information, contained in the system's cause-effect structure (if there is no mechanism for it, it does not matter to the system).

Yet, the system itself does not 'have' this intrinsic information. Just by 'processing' information, a system cannot evaluate its own constraints. This is simply because a system cannot, at the same time, have information about itself in its current state and also other possible states. Any memory the system has about its past states has to be physically instantiated in its current cause-effect structure. While a system can have mechanisms that, by being in their current state, constrain other parts of the system, these mechanisms cannot 'know' what their inputs mean[6]. In the same sense, a system of mechanisms in its current state does not 'know' about its cause-effect structure; instead, the cause-effect structure specifies what it means to *be* the system in a particular state[7]. Intrinsic meaning thus cannot arise from 'knowing', it must arise from 'being'.

What does it mean to 'be' a system, as opposed to an assembly of interacting elements, defined by an extrinsic observer? When can a system of mechanisms be considered an autonomous agent separate from its environment?

## IV. To be or not to be integrated

Living systems, or agents, more generally, are, by definition, open systems that dynamically and materially interact with their environment. For this reason, physics, as a set of mathematical laws governing dynamical evolution, does not distinguish between an agent and its environment. When a subsystem within a larger system is characterized by physical, biological, or informational means, its boundaries are typically taken for granted (see also [16]).

Let us return to the Markov Brains shown in Fig. 2, which evolved perfect solutions in the c1-a3 environment. Comparing the two network architectures, the Markov Brain in Fig. 2A has only feedforward connections between elements, while the hidden elements in Fig. 2B feedback to each other. Both Markov Brains 'process' information in the sense that they receive signals from the environment and react to these signals. However, the hidden elements in Fig. 2B constrain each other, above a background of external inputs, and thus from an *integrated* system of mechanisms.

Whether and to what extent a set of elements is integrated can be determined from its cause-effect structure, using the theoretical framework of integrated information theory (IIT) [13]. A subsystem of mechanisms has integrated information $\Phi > 0$, if all of its parts constrain, and are being constrained by, other parts of the system. Every part must be a difference that makes a difference within the subsystem. Roughly, $\Phi$ quantifies the minimal intrinsic information that is lost if the subsystem is partitioned in any way. An integrated subsystem with $\Phi > 0$ has a certain amount of causal autonomy from its environment[8]. Maxima of $\Phi$ define where intrinsic causal borders emerge [17,18]. A set of elements thus forms a causally autonomous entity if its mechanisms give rise to a cause-effect structure with maximal $\Phi$, compared to smaller or larger overlapping sets of elements. Such a maximally integrated set of elements forms a unitary whole (it is 'one' as opposed to 'many') with intrinsic, self-defined causal borders, above a background of external interactions. By contrast, systems whose elements are connected in a purely feedforward manner have $\Phi = 0$: there is at least one part of the system that remains unconstrained by the rest. From the intrinsic perspective, then, there is no unified system, even though an external observer can treat it as one.

So far, we have considered the entire Markov Brain, including sensors, hidden elements, and motors, as the system of interest. However, the sensors only receive input from the environment, not from other elements within the system, and the motors do not output to other system elements. The whole Markov Brain is not an integrated system, and thus not an autonomous system, separate from its environment. Leaving aside the animat's 'retina' (sensors) and 'motor neurons' (motors), inside the Markov Brain in Fig. 2B, we find a minimal entity with $\Phi > 0$ and self-defined causal borders—a 'brain' within the Markov Brain. By contrast, all there is, in the case of Fig. 2A, is a cascade of switches, and any border demarcating a particular set of elements would be arbitrary.

---

[5] By contrast to the uniform, perturbed distribution, the stationary, observed distribution of system $Z$ entails correlations due to the system's network structure which may occlude or exaggerate the causal constraints of the mechanism itself.

[6] Take a neuron that activates, for example, every time a picture of the actress Jennifer Aniston is shown [22]. All it receives as inputs is quasi-binary electrical signals from other neurons. The meaning "Jennifer Aniston" is not in the message to this neuron, or any other neuron.

[7] For example, an AND logic gate receiving 2 inputs is what it is, because it switches ON if and only if both inputs were ON. An AND gate in state ON thus constrains the past states of its input to be ON.

[8] This notion of causal autonomy applies to deterministic and probabilistic systems, to the extent that their elements constrain each other, above other background inputs, e.g. from the sensors.

Dynamically and functionally the two Markov Brains are very similar. However, one is an integrated, causally autonomous entity, while the other is just a set of elements performing a function. Note again that the two systems are equally 'intelligent' (if we define intelligence as task fitness). Both solve the task perfectly. Yet, from the intrinsic perspective being a causally autonomous entity makes all the difference (see here [13,19]). But is there a practical advantage?

## V. Advantages of being integrated

The cause-effect structure of a causally autonomous entity describes what it means to be that entity from its own intrinsic perspective. Each of the entity's mechanisms, in its current state, corresponds to a distinction within the entity. Being an entity for which 'light' is different from 'dark', for example, requires that the system itself, its cause-effect structure, must be different when it 'sees' light, compared to when it 'sees' dark. In this view, intrinsic meaning might be created by the specific way in which the mechanisms of an integrated entity constrain its own past and future states, and by their relations to other mechanisms within the entity.

The animat 'brain' in Fig. 2B, constituted of the 2 hidden elements, has at most 3 mechanisms (each element, and also both elements together, if they irreducibly constrain the system). At best, these mechanisms could specify that "something is probably this way, not that way", and "same" or "different". Will more complex environments lead to the evolution of more complex autonomous agents?

In the simple c1-a3 environment, animats with integrated brains do not seem to have an advantage over feedforward architectures. Out of the 13 strains of animats that reached perfect fitness, about half developed architectures with recurrent connections (6/13) [5]. However, in a more difficult block-catching environment, which required more internal memory ("catch size 3 and 6, avoid size 4 and 5"), the same type of animats developed more integrated architectures with higher $\Phi$, and more mechanisms (one example architecture is shown in Fig. 1B). The more complex the environment, the more evolution seems to favor integrated structures.

In theory, and more so for artificial systems, being an autonomous entity is not a requirement for intelligent behavior. Any task could, in principle, be solved by a feedforward architecture given an arbitrary number of elements and updates. Nevertheless, in complex, changing environments, with a rich causal structure, where resources are limited and survival requires many mechanisms, integrated agents seem to have an evolutionary advantage [5,20]. Under these conditions, integrated systems are more economical in terms of elements and connections, and more flexible than functionally equivalent systems with a purely feedforward architecture. Evolution should also ensure that the intrinsic cause-effect structure of an autonomous agent 'matches' the causal structure of its environment [21].

From the animats, it is still a long way towards agents with intrinsic goals and intentions. What kind of cause-effect structure is required to experience goals, and which environmental conditions could favor its evolution, remains to be determined. Integrated information theory offers a quantitative framework to address these questions.

## VI. Conclusion

Evolution did produce autonomous agents. We experience this first hand. We are also entities with the right kind of cause-effect structure to experience goals and intentions. To us, the animats appear to be agents that behave with intention. However, the reason for this lies within ourselves, not within the animats. Some of the animats even lack the conditions to be separate causal entities from their environment. Yet, observing their behavior affects *our* intrinsic mechanisms. For this reason, describing certain types of directed behaviors as goals, in the extrinsic sense, is most likely useful to us from an evolutionary perspective. While we cannot infer agency from observing apparent goal-directed behavior, by the principle of sufficient reason, something must cause this behavior (if we see an antelope running away, maybe there is a lion). On a grander scale, descriptions in terms of goals and intentions can hint at hidden gradients and selection processes in nature, and inspire new physical models.

For determining agency and intrinsic meaning in other systems, biological or not, correlations between external and internal states have proven inadequate. Being a causally autonomous entity from the intrinsic perspective requires an *integrated* cause-effect structure; merely 'processing' information does not suffice. Intrinsic goals certainly require an enormous amount of mechanisms. Finally, when physics is reduced to a description of mathematical laws that determine dynamical evolution, there seems to be no place for causality. Yet, a (counterfactual) notion of causation may be fundamental to identify agents and distinguish them from their environment.

### Acknowledgements

I thank Giulio Tononi for his continuing support and comments on this essay, and William Marshall, Graham Findlay, and Gabriel Heck for reading this essay and providing helpful comments. L.A. receives funding from the Templeton World Charities Foundation (Grant #TWCF0196).